# Charge Ordering in Amorphous WO$_x$ Films

Yakov Kopelevich and Robson R. da Silva

Instituto de Física "Gleb Wataghin", Universidade Estadual de Campinas, Unicamp 13083-970, Campinas, São Paulo, Brasil

Aline Rougier

Laboratoire de Réactivité et Chimie des Solides, Université de Picardie Jules Verne, Amiens, 80039, France

Igor A. Luk´yanchuk

Laboratoire de Physique de Matière Condensée, Université de Picardie Jules Verne, Amiens, 80039, France, and L. D. Landau Institute for Theoretical Physics, Moscow, Russia

ABSTRACT

We report on the observation of highly anisotropic viscous electronic conducting phase in amorphous WO$_{1.55}$ films that occurs below a current (I)- and frequency (f)-dependent temperature T$^*$(I, f). At T < T$^*$(I, f) the rotational symmetry of randomly disordered electronic background is broken leading to the appearance of mutually perpendicular metallic- and insulating-like states. A rich dynamic behavior of the electronic matter occurring at T < T$^*$(I, f) provides evidence for an interplay between pinning effects and electron-electron interactions. The results strongly suggest a dynamic crystallization of the disordered electronic matter, viz. formation of sliding Wigner crystal, as well as the occurrence of quantum smectic or stripe phase in the pinning-dominated regime.

Transition-metal oxides (TMO) attract a broad scientific interest due to their extraordinary physical properties such as, e.g., high temperature superconductivity [1] and colossal magnetoresistance [2]. The interplay between different ground states of the electronic matter is one of the central issues of the TMO physics. In particular, a profound relationship between the above phenomena and quantum dynamics of quasi-one-dimensional electronic stripes is expected [1-3]. The theory of quantum electronic liquid crystals [3, 4] predicts the occurrence of either superconducting or non-Fermi-liquid ground states associated with dynamics of charge-ordered phases. Experiments revealed the charge ordering in various TMO, indeed [1, 2]. However, all the studies were performed on crystalline TMO compounds, where the atomic (ionic) long-range order inevitably affects, or perhaps governs the stripe formation.

In this Letter we report on the observation of anisotropic charge-ordered electronic phase in amorphous $WO_x$ (x ~1.55) films, characterized by broken rotation symmetry below the temperature $T^*$(I, f) of the order of ~ 10 K that is current (I) and frequency (f) dependent. Measurements of anisotropic low-frequency resistance and current-voltage (I-V) characteristics at T< $T^*$ revealed generic features of moving many-body interacting systems such as Wigner crystals (WC), Wigner glasses or smectics [5, 6], electronic stripe phases [7], and Abrikosov vortex matter [8-10].

Tungsten trioxides $WO_3$ can be described as a network of corner-linked $WO_6$ octahedra resulting in the occurrence of several crystallographic phases ranging from triclinic or monoclinic phases to the tetragonal one [11]. Tungsten dioxide $WO_2$ possesses even more complicated monoclinically distorted rutile structure [12]. The stoichiometric $WO_3$ is an insulator because all six ($6s^2 6d^4$) valence electrons of tungsten $W^{6+}$ are captured by surrounded oxygen $O^{2-}$. Experimentally, the low-temperature activation energy for the electron conduction $\Delta$ = 0.3 eV or 0.23 eV has been reported for $WO_3$ amorphous films [13] and the bulk crystal [14], respectively. Doping of $WO_3$ with electropositive donors M (H, Na, K, Rb, Cs, In, Ta) leads to a metallic or even superconducting $M_xWO_3$ compounds [15], typically having superconducting transition temperature $T_c$ < 4 K. Noting, however, observations of possible high temperature surface superconductivity in Na-doped $WO_3$ with $T_c$ = 91 K [16]. On the other hand, the superconductivity below $T_c$ = 3 K was observed in twin walls of reduced $WO_{3-\delta}$ crystals [17].

Here we explored the possibility of a conductivity enhancement in reduced amorphous $WO_x$ films. The studied $WO_x$ films were grown on glass substrates from $WO_3$ target under vacuum P ≈ (5-8) $10^{-4}$ Pa or oxygen pressure $PO_2$ ≈ 1 Pa at room temperature by pulsed laser deposition using a KrF excimer laser (Lambda Physik, Compex 102, $\lambda$ = 248 nm) with a laser pulse energy of 180 mJ and a rate frequency of 3 Hz [18]. The film thickness 400 nm < d < 500 nm was measured by a DEKTAK profilometer. The Rutherford Backscattering Spectrocopy (RBS) analysis revealed $WO_{1.55}$ and $WO_{2.33}$ homogeneous films obtained under vacuum ($WO_{1.55}$) and 1 Pa oxygen pressure ($WO_{2.33}$). Featureless x-ray diffraction patterns measured for all studied films provide evidence for the film amorphization.

Van der Pauw - type measurements of the longitudinal R(T, f, I, B) resistance vs. temperature (T), frequency (f), applied electrical current (I) and magnetic field (B) have been performed on several amorphous $WO_x$ films for 2 K ≤ T ≤ 300 K, 1 Hz ≤ f ≤ 1 kHz, 10 µA ≤ I ≤ 1 mA, and 0 ≤ B ≤ 9 T by means of PPMS (Quantum Design) and Janis 9 T-magnet He-cryostats. For the measurements, four Pt point-like contacts were patterned at corners of square- or rectangular-shaped films. The resistances $R_{14,23} = V_{23}/I_{14}$ and $R_{12,34} = V_{34}/I_{12}$ were measured as shown in the inset of Fig. 3 (a) and Fig. 3 (b) respectively.

All studied $WO_{1.55}$ films with dimensions 0.72 x 0.38 x 0.0004 mm$^3$ (samples F1 and F3), and 0.84 x 0.8 x 0.0004 mm$^3$ (sample F4) have the room-temperature resistivity $\rho$ (300 K) = 5.0 ± 0.2 m$\Omega$cm. Dimensions and resistivity of the $WO_{2.33}$ film, labeled here as F2 are 0.8 x 0.75 x 0.00049 mm$^3$ and $\rho$(300 K) = 45 ± 5 m$\Omega$cm, respectively.

Figure 1 presents $R(T) \equiv R_{14,23}(T)$ measured for the sample F1 with I = 0.1 mA, f = 1 Hz, for B = 0 and B = 5 T. The prominent feature of given in Fig. 1 R(T) curves is the resistance drop that occurs at $T < T_{max}(I) \equiv T^*(I) \sim 6.5$ K, insensitive to the applied magnetic field [19]. Fig. 1 illustrates also that for $T > T_{max}$, R(T) can be well described by the equation

$$R(T) = \frac{1}{A + BT^\alpha}, \qquad (1)$$

where $\alpha = 0.85$. Similar R vs. T behavior with $\alpha = 1.00 \pm 0.15$ was found for all measured $WO_{1.55}$ films. On the other hand, R(T) obtained for the film F2, $WO_{2.33}$ diverges (see inset in Fig. 1) with the temperature lowering as

$$R(T) = R_M \exp[(T_M/T)^{1/3}], \qquad (2)$$

or

$$R(T) = R_{ES} \exp[(T_{ES}/T)^{1/2}], \qquad (3)$$

i. e. R(T) obeys the Mott´s (M) law for variable range hopping (VRH) of localized electrons in two-dimensional (2D) systems [20, 21] or Efros-Shklovskii (ES) law that implies a Coulomb gap existence in the VRH regime [22].

Figure 2 shows R(T) obtained for the film F1 with various measuring currents and f = 1 Hz. Fig. 2 demonstrates that R(I, T = const) is a decreasing function of I, as well as that $T_{max}(I)$ increases with the applied current. The best fitting to the data (dashed line in the inset of Fig. 2) gives $I \sim (T_{max})^\beta$ where $\beta = 2.6$.

In Figures 3 - 5 we present the results that further characterize the intriguing low-temperature resistance behavior revealed in $WO_{1.55}$ films.

Figure 3 (a, b) presents low-temperature portions of $R_{12,34}(T)$ and $R_{14,23}(T)$ obtained for the film F3 at various frequencies and measuring current $I = 10$ µA. As can be seen form this figure, for temperatures below $\sim 10$ K, $R_{12,34}(T)$ and $R_{14,23}(T)$ demonstrate a pronounced frequency dependence. At low enough frequencies (f = 10 Hz), and down to the lowest measuring temperature T = 2 K, $dR_{14,23}/dT < 0$ and $dR_{12,34}/dT > 0$, such that the ratio $R_{14,23}/R_{12,34}$ diverges (not shown) with temperature lowering according to the formula $R_h/R_e = A_1 + B_1 \exp(C/T^{1/2})$, with $A_1 = 1.5$, $B_1 = 0.003$, and $C = 12[K^{1/2}]$. This $R_h/R_e$ vs. T behavior suggests the formation of an ideal metal in the "easy" (more conductive) direction or/and an insulator - in the "hard" (less conductive) direction ground states ($A_1 > 1$ originates from the sample geometrical factor). At the same time, for our highest measuring frequency (f = 1 kHz) the resistance drops below $T_{max}$ by two orders of magnitude, independently of the applied current direction. As the low inset in Fig. 3(b) illustrates, $f \sim (T_{max})^\gamma$ where $\gamma = 2.7$. Noting, $f(T_{max})$ can be approximated by the Arrhenius-type law expected for a simple thermally-activated process only at T > 4 K.

Thus, Fig. 2 and Fig. 3 provide evidence that the increase of either current amplitude or its frequency reduces the resistance and its anisotropy at $T < T_{max}(I, f)$. Because $I \sim (T_{max})^\beta$ and $f \sim (T_{max})^\gamma$ correspond to the same power law ($\beta \approx \gamma$), it is reasonable to assume a similar physical mechanism behind of the non-linear and frequency-dependent resistance.

Figure 4 exemplifies current-voltage I(V) characteristics measured at $T \ll T_{max}$ and $f = 60$ Hz for the film F4 with current flowing in both "hard" (curve 1) and "easy" (curve 2) directions. The I(V) measured for the film F3 in "hard" direction (curve 3) is also shown. Inset in Fig. 4 gives differential conductance $G \equiv dI/dV$ vs. V dependencies derived from I(V) curves presented in Fig. 4.

Shown in Fig. 4, I(V) and G(V) dependencies resemble very much the curves measured for various driven macroscopic elastic objects interacting with a quenched disorder [5-10], such as e. g. Abrikosov vortex lattices and charge-ordered electronic systems, where an interplay between inter-particle (vortices or electrons) and particle-quenched disorder interactions plays a crucial role. In these systems, an effective strength of the interactions can be controlled by the applied driving force. At low drives the particle motion is dominated by the quenched disorder (pinning) effects, leading to the plastic (channel-like) motion of some regions of the elastic medium with respect to other temporarily pinned regions. With increasing the driving force, (re)ordering or dynamic crystallization of the elastic medium takes place [7, 10]. Whereas our results revealed no signature for the superconductivity and hence vortices, Wigner crystallization, electronic stripe or other many-electron phase formation are plausible scenarios. According to theoretical models [7], the peak in G(V) at $V \equiv V_p$, see the inset in Fig. 4, is expected when the disordered electronic matter becomes dynamically more ordered, forming Wigner crystal or other ordered electronic phases. Because $V_p$ in the film F3 exceeds that in the F4 sample, one concludes that pinning in the sample F3 is stronger. Considering a combined effect of quenched disorder and electron-electron interactions, R vs. f behavior shown in Fig. 3 can be accounted for by the frequency-induced depinning of the elastic electronic media, in a close analogy with the depinning occurring in the vortex matter [23, 24]. The frequency-induced resistance drop is also expected for pinned WC at $f < f_p$, where $f_p$ is the characteristic de(pinning) frequency, see e. g. Ref. [25]. In the opposite limit, i. e. for $f > f_p$, R is an increasing function of the frequency [25]. Figure 5 illustrates such an "inverse" R(f) dependence measured for the film F4 which possesses a weaker pinning. The inset in Fig. 5 presents $R_h$(T) for the same sample F4 measured with f = 1 kHz and 20 µA ≤ I ≤ 400 µA. We stress, that while in the low-frequency, large-current limit, R grows monotonically with the temperature increase, the non-monotonic R(T) takes place for low currents and high enough frequencies (Fig. 5). Clearly, the minima in R(T) develop at higher resistance level, corresponding to an incoherent state where a competition between pinning effects and electron-electron interactions should be of a particular importance. However, as temperature increases up to T ≈ $T_{min}$, see Fig. 5, inter-electron interactions dominate, and a more coherent state characterized by the lower resistance is formed. At yet higher temperatures, the crystal melts and the resistance increases once again. If our interpretation is correct, this "dip effect" may well be related to the temperature-induced crystallization also reported for vortex [26-28] and polymeric [29] systems.

To get more insight on the origin of the possible collective electronic state in our films, we note that R(T) described by Eq. (1), see Fig. 1, is expected for systems in the vicinity of a metal-insulator transition (MIT), characterized by the incoherent electronic transport [30]. The resistance drop at T < $T_{max}$ indicates the opening of an additional coherent channel. Taking $k_F l \sim 1$ at MIT and $\sigma(T_{max}) = (e^2/3\pi^2\hbar)k_F^2 l \sim 100$ $\Omega^{-1}$cm$^{-1}$, one gets $k_F \sim 10^7$ cm$^{-1}$ that translates to the carrier density n = $(k_F)^3/3\pi^2 \sim 10^{18}$ cm$^{-3}$. For such n, Wigner crystallization is expected to occur at $T_{WC}$ ≤ 10 K [31] which is not inconsistent with our observations. The occurrence of WC in $WO_{1.55}$ films is also supported by the striking similarity of our results with observations of WC in other systems. Thus, the resistance drop at $T \leq T_{WC}$, the peak in G(V) and its similar frequency dependence, all has been observed in 2D WC formed on a surface of superfluid $^4$He [32]. Besides, our data also suggest the opening of a Coulomb gap in the sample F2 ($WO_{2.33}$), see inset in Fig. 1, that can be considered as a precursor of WC [33]; the temperature crossover from Efros-Shklovskii to Mott hopping regime [33] takes place at T ~ 10 K.

It is interesting to note that an anisotropic conducting state with a broken rotation symmetry, reported here for $WO_x$ films, has also been seen in 2D electron systems in quantized magnetic field [34], pointing out to a certain universality in the behavior of strongly correlated electrons.

On the other hand, we would like to emphasize the relevance of competing long-range Coulomb repulsive and short-range attractive interactions [7, 35, 36] for the occurrence of various collective electronic phases. In the case of $WO_x$ the short-range attraction can originate from a strong electron-phonon (e-ph) interaction [37, 38]. This fact is of a particular interest in the light of recent experiments performed on superconducting high-$T_c$ copper oxides [39] that revealed the coupling between e-ph interaction and charge inhomogenities.

Summing up, we observed the anomalous metallic phase in amorphous $WO_{1.55}$ films that demonstrates characteristic features of moving many-body interacting electron systems. Further work is needed to clarify a microscopic origin of competing interactions in the tungsten oxide films.

This work was supported by FAPESP, CNPq, CAPES, and COFECUB.

FIGURE CAPTIONS

Fig. 1. Temperature dependence of the resistance R(T) measured for sample F1 with I = 0.1 mA, f = 1 Hz for zero (B = 0) and applied magnetic field B = 5 T. Dashed line corresponds to Eq. (1) with parameters A = 0.0346 $\Omega^{-1}$, B = 0.000525 $\Omega^{-1}K^{-0.85}$ ; charge ordering takes place at T ≤ $T_{max}$ (see text). The inset shows R(T) measured for the film F2 at f = 1 Hz, B = 0; solid and dashed lines are obtained from Eq. (2) and Eq. (3) respectively; $T_M$ = 400 K, $R_M$ = 25 Ω, $T_{ES}$ = 40 K, $R_{ES}$ = 110 Ω.

Fig. 2. Temperature dependencies of the resistance R(T) obtained for film F1 using various measuring currents, f = 1 Hz, and B = 0. $T_{max}$(I) increases with the applied current according to the power law I ~ $(T_{max})^{2.6}$ (see inset).

Fig. 3. Temperature dependencies of resistances measured in "easy" $R_e$ = $V_{34}/I_{12}$, inset in (a), and "hard" $R_h$ = $V_{23}/I_{14}$, upper inset in (b), directions for B = 0, $I_{12}$ = $I_{14}$ = 10 µA and frequencies (a) f = 10, 100, 200, 500, 1000 Hz; (b) f = 10, 25, 50, 100, 200, 500, 1000 Hz, from top to bottom. Lower inset in (b) shows variation of $T_{max}$ with frequency; dotted line corresponds to the power law f ~ $(T_{max})^{2.7}$.

Fig. 4. Current-voltage I(V) characteristics measured at T = 2 K, B =0 and f = 60 Hz for the film F4 with current flowing in both "hard" (curve 1) and "easy" (curve 2) directions; curve 3 is I(V) measured for the film F3 ( T = 5 K, B =0, f = 60 Hz) in "hard" direction. The inset shows differential conductance $G \equiv dI/dV$ vs. V dependencies (1, 2, and 3) derived from the corresponding I(V) curves; $V_p$ marks the maximum in G(V).

Fig. 5. Temperature dependencies of resistance $R_h(T)$ in "hard" direction for sample F4 at B = 0, I = 10 µA and frequencies f = 50, 100, 200, 300, 400, 600, 1000 Hz (from bottom to the top). The inset shows $R_h(T)$ measured for the same sample at B = 0, f = 1 kHz, and currents I = 20, 30, 40, 60, 70, 80, 90, 100, 150, 200, 300, 400 µA (from top to the bottom). Arrows mark $T_{max}$ and $T_{min}$ (see text).

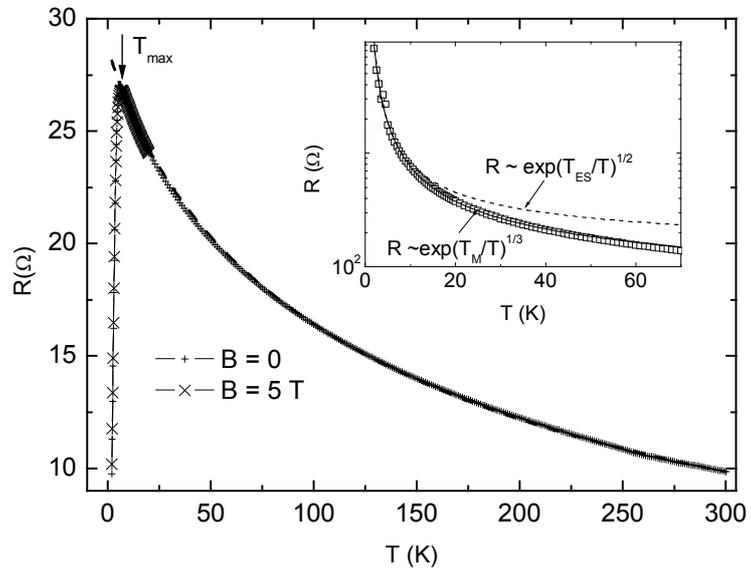

Fig. 1

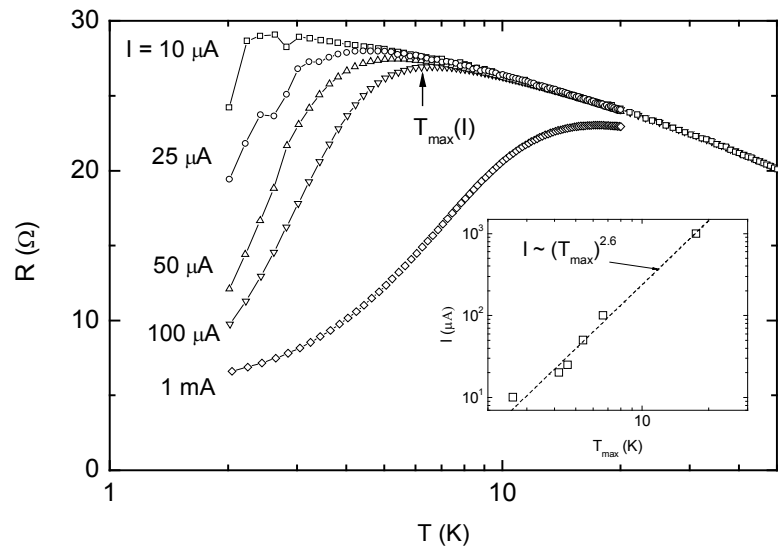

Fig. 2

Fig. 3

(a) [plot of $R_e$ ($\Omega$) vs $T$ (K) with inset showing sample schematic: current $I$ from contact 1 to 2, voltage $V$ measured between contacts 4 and 3]

(b) [plot of $R_h$ ($\Omega$) vs $T$ (K) with inset showing sample schematic: current $I$ from contact 1 to 4, voltage $V$ measured between contacts 2 and 3; second inset: log-log plot of $f$ (Hz) vs $T_{max}$ (K) showing $f \sim (T_{max})^{2.7}$]

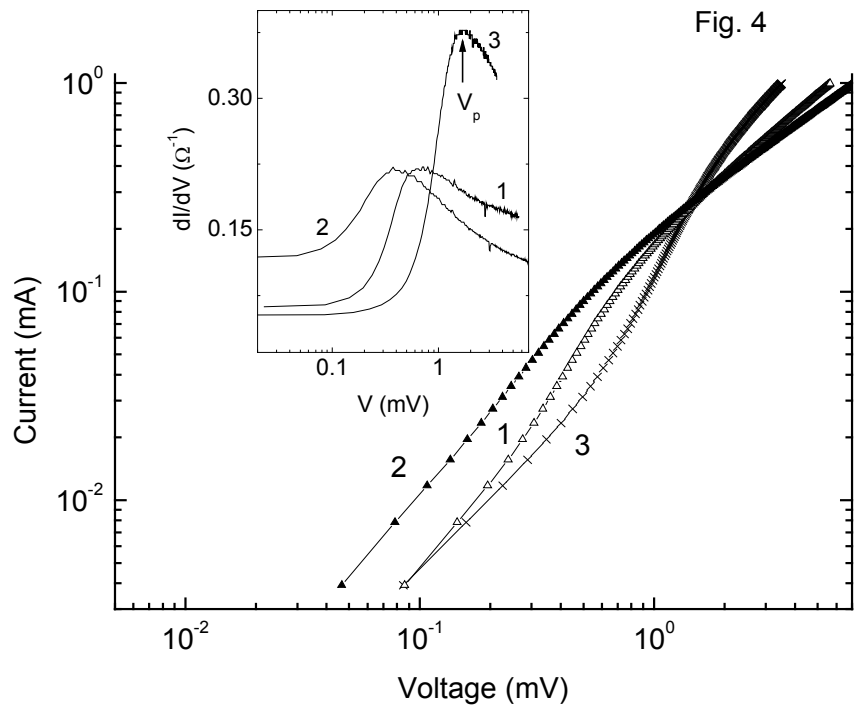

Fig. 4

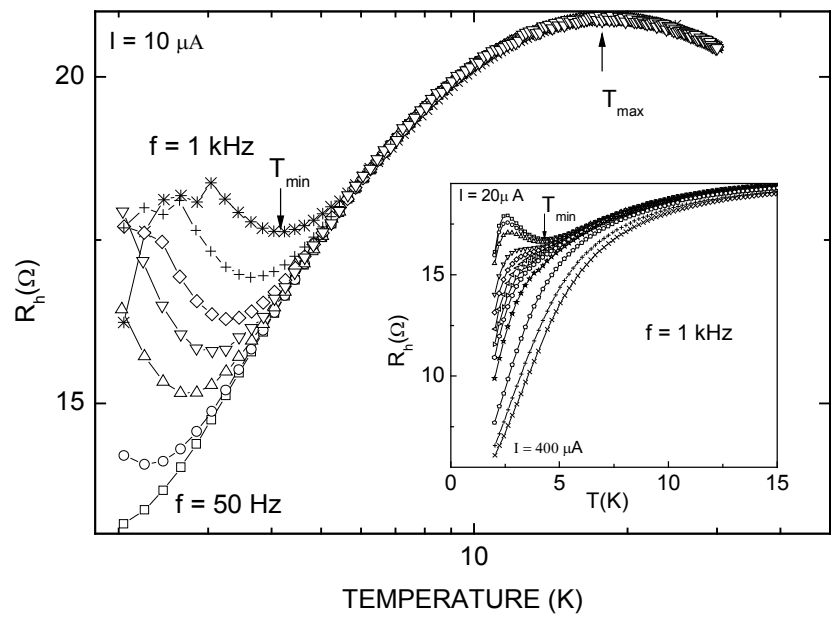

Fig. 5